\documentclass[fleqn,twoside]{article}
\usepackage{espcrc2}

\usepackage{graphicx}
\newcommand{\etal}{\emph{et al.}}

\title{Vorticity, phase stiffness and the cuprate phase diagram}

\author{N. P. Ong$^a$ and Yayu Wang
\address{
Department of Physics, Princeton University, Princeton, N.J. 08544, U. S. A.}
        \thanks{The research reported is a collaboration with Z. A. Xu, S. Uchida, S. %%@
Ono, Yoichi Ando, G. Gu, Y. Onose, Y. Tokura, D. A. Bonn, R. Liang, and W. N. Hardy.  We %%@
acknowledge support from the U. S. National Science Foundation (NSF), the U. S. Office %%@
of Naval Research, and the New Energy and Industrial Technology Developmental %%@
Organization of Japan. Some of the experiments were performed at the National High %%@
Magnetic Field Lab. (NHMFL), a facility supported by NSF and the State of Florida.} }

\begin{document}

\begin{abstract}
We review results obtained from vortex-Nernst experiments in cuprates.  Evidence for a %%@
loss of phase coherence at the Meissner transition $T_{c0}$ is derived from vortex-like %%@
excitations that persist to high temperature $T$.  Below $T_{c0}$, the Nersnt signal %%@
provides a determination of the upper critical field $H_{c2}$ vs. doping $x$.  %%@
Implications for the cuprate phase diagram are discussed.
\vspace{1pc}
\end{abstract}

%vortex \sep Nernst effect \sep critical field \sep cuprates
%\PACS 74.40.+k \sep 72.15.Jf \sep 74.25.Fy \sep 74.72.-h

\maketitle

The vortex-Nernst effect is a highly sensitive probe for detecting vortex motion in a %%@
type II superconductor~\cite{Kim}.  In the past 3 years, we have used it to map out the %%@
region in the field-temperature ($H$-$T$) plane in which vorticity may be %%@
observed~\cite{Xu,Wang1,Wang2,Wang3}.  The results provide a fresh perspective on the %%@
cuprate phase diagram which we sketch here.   When a superconductor (in the %%@
vortex-liquid state) is exposed to a weak gradient $-\nabla T|| {\bf \hat{x}}$ in a %%@
field ${\bf H||\hat{z}}$, vortices diffuse down the gradient with velocity $\bf %%@
v||\hat{x}$.  As each vortex \emph{core} crosses the line between a pair of transverse %%@
voltage electrodes, the $2\pi$ phase slip of the condensate phase leads to a Josephson %%@
$E$ field given by $\bf E = B\times v$.  The Nernst signal is defined as $e_y = %%@
E_y/|\nabla T|$.  In cuprates, Nernst experiments were initially conducted on %%@
optimally-doped samples~\cite{Hagen}.

\begin{figure}[h]		% Fig 1
\includegraphics[width=6cm]{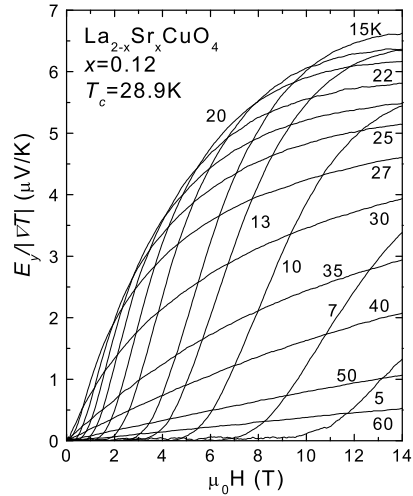}
\caption{\label{eyLSCO} Field dependence of Nernst signal $e_y$ in LSCO ($x = 0.12$) at %%@
$T$ above and below $T_{c0}$.  From Ref.~\cite{Wang1}.
}
\end{figure}

In extending the experiments to underdoped $\rm La_{2-x}Sr_xCuO_4$ (LSCO), Xu %%@
\etal~\cite{Xu} observed that $e_y$ persists to temperatures 50-100 K above $T_{c0}$.  %%@
Figure \ref{eyLSCO} shows several curves of $e_y$ in underdoped LSCO ($x = 0.12$).  %%@
Below the zero-field transition temperature $T_{c0}\simeq$ 29 K, $e_y(T,H)$ is initially %%@
zero until the melting field line $H_m(T)$ is exceeded (for e.g., at 5 T in the curve at %%@
10 K).  In the liquid state, $e_y$ climbs rapidly to attain a broad maximum near 14 T.  %%@
As we warm to $T_{c0}$, we find that the maximum in $e_y$ (curve at 30 K) is not much %%@
smaller than the low-$T$ maxima.  If $e_y$ is linear in $H$ in weak fields (above %%@
$T_{c0}$), we may define the Nernst coefficient as $\nu = e_y/B$ ($B\rightarrow 0$).  %%@
Above $T_{c0}$, $\nu$ falls slowly and remains observable to $\sim$130 %%@
K~\cite{Xu,Wang1,Capan}.

\begin{figure} [h]		% Fig 2
\includegraphics[width=6cm]{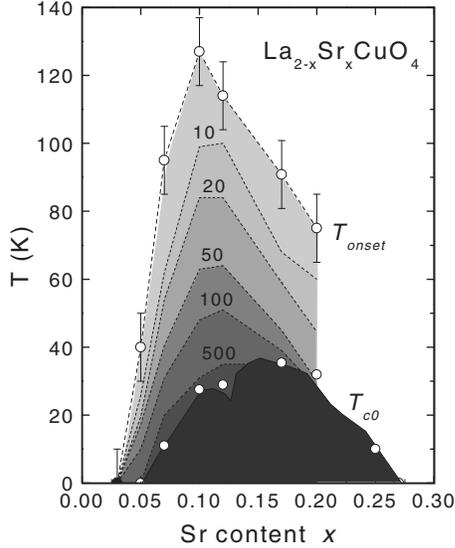}
\caption{\label{vortexphase} Phase diagram showing contours of the Nernst coefficient %%@
$\nu$ above the critical transition line $T_{c0}$ vs. $x$ in LSCO.  The number at each %%@
contour line is the value of $\nu$ in $nV/KT$.  Note that all contours and the onset %%@
line $T_{onset}$ peak near $x = 0.10$.   From Ref.~\cite{Wang1}.
}
\end{figure}
\begin{figure} [h]			% Fig3
\includegraphics[width=6cm]{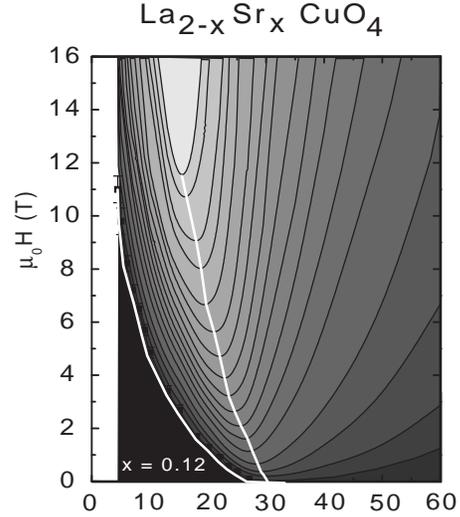} 
\caption{\label{contour} Contour plot of the Nernst signal $e_y(T,H)$ in the $H$-$T$ %%@
plane for LSCO $x$ = 0.12.  Light grey indicates regions with largest value of $e_y$, %%@
while black indicates $e_y = 0$ (vortex solid).  The melting field $H_m(T)$ (ridge field %%@
$H^*(T)$) is the lower (upper) white curve.  
}
\end{figure}
The findings of Xu \etal~\cite{Xu} imply that vortex-like excitations exist above %%@
$T_{c0}$ high into the pseudogap state.   What is the onset temperature $T_{onset}$?  At %%@
high $T$, where the vortex-Nernst coefficient $\nu$ becomes comparable to that of the %%@
carriers, it is necessary to measure the hole thermopower and Hall angle to isolate the %%@
vortex signal~\cite{Wang1}.  The derived phase diagram for LSCO shows that $T_{onset}$ %%@
lies high above $T_{c0}$ (Fig. \ref{vortexphase}).  A notable feature is the prominent %%@
maximum of $T_{onset}$ and all the contours at $x\sim 0.1$ (instead of 0.17).  These %%@
results provide strong evidence that, over a large part of the phase diagram, %%@
significant condensate strength exists above $T_{c0}$.  This raises the possibility that %%@
the line $T_{c0}$ vs. $x$ measured by the Meissner effect actually corresponds to the %%@
loss of long-range phase coherence instead of the vanishing of the superconducting %%@
complex order parameter $\hat{\psi}({\bf r})$~\cite{Kivelson,Corson}.  

Below $T_{c0}$, the dependence of $e_y(T,H)$ on $T$ and $H$ changes in a characteristic %%@
way as a function of doping.  An effective way to provide a broad overview is display %%@
$e_y(T,H)$ as a contour map in the $H$-$T$ plane~\cite{Wang2}.  In Fig. \ref{contour} %%@
(for LSCO, $x = 0.12$), $e_y$ attains its largest value in the light areas, while it is %%@
zero in the black areas (in the vortex-solid phase).  As the melting line $H_m(T)$ line %%@
is crossed in a fixed-$H$ scan, the signal rises steeply to a maximum before decreasing %%@
slowly on the high-$T$ side.  The locus of the maxima defines a `ridge' field $H^*(T)$.  %%@
We stress that, in the contour map, no crossover line or phase boundary separates the %%@
vortex liquid phase from a putative normal state above $T_{c0}$.  

In low-$T_c$ superconductors, e.g. 2$H$-NbSe$_2$, the upper critical field line %%@
$H_{c2}(T)$ unambiguously separates the Abrikosov state from the normal state.  %%@
Moreover, $H_{c2}$ approaches zero linearly as $H_{c2}\sim (T_{c0}-T)$.  Where is %%@
$H_{c2}$ in the cuprates?  The received wisdom seems to be that (i) $H_{c2}$ in cuprates %%@
is completely obliterated by strong fluctuations and not observable, or (ii) the `real' %%@
$H_{c2}$ line should be identified with $H_m(T)$ since this is the line at which %%@
superfluidity vanishes.  Our experiments do not support either view.

\begin{figure}[h]			% Fig 4
\includegraphics[width=8cm]{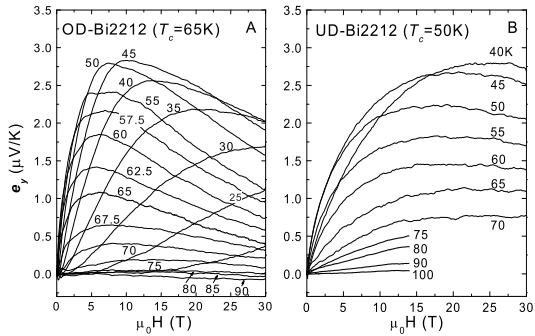}
\caption{\label{Bi2212comp}  Curves of $e_y$ vs. $H$ in overdoped (Panel A) and %%@
underdoped (Balnel B) Bi 2212.  The Nernst signal peaks at 5-10 T in the overdoped %%@
sample but at larger $H$ (15-30 T) in the underdoped.  Bold lines are curves taken at %%@
$T_{c0}$ in both samples.  From Ref.~\cite{Wang3}
}
\end{figure}

\begin{figure}[h]			% Fig 5
\includegraphics[width=7cm]{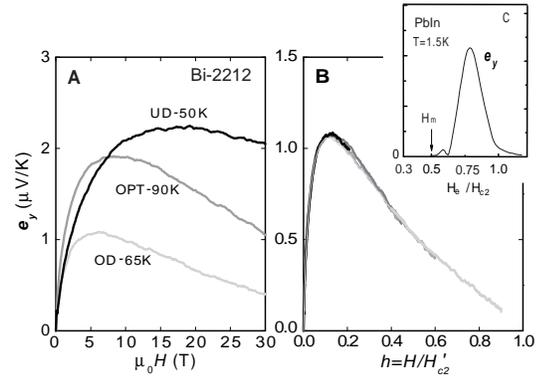}
\caption{\label{scale}  Panel A: Curves of $e_y$ taken at $T_{c0}$ in overdoped (OD), %%@
optimum-doped (OPT) and underdoped (UD) Bi 2212.  Panel B shows the collapse of the 3 %%@
curves onto the template curve derived from $\rm Bi_2Sr_{2-y}La_yCuO_6$ ($y = 0.4$) when %%@
plotted versus the reduced field $h = H/H_{c2}$.  Panel C shows $e_y$ vs. $H$ in PbIn %%@
(from an Ettingshausen experiment~\cite{Vidal}).  Modified from Ref.~\cite{Wang3}
}
\end{figure}
In our quest for $H_{c2}$, we have extended measurements to 30 T (later to 45 T) at %%@
NHMFL.  The higher fields immediately revealed that, in every sample, the curve of $e_y$ %%@
vs. $H$ invariably has a `tent' profile.  To understand its significance, we examined %%@
how $e_y$ behaves in thin-film PbIn.  There, $e_y$ (derived from the Ettingshausen %%@
effect~\cite{Vidal}) increases rapidly when the vortex lattice is depinned, then rises %%@
to a sharp maximum before falling to zero linearly with the difference field $H_{c2}-H$ %%@
(Fig. \ref{scale}c).  The decrease reflects the field suppression of the condensate %%@
amplitude. 

In the cuprates, the contour plots provide a road map for the field needed to get over %%@
the ridge field $H^*(T)$.  Beyond $H^*(T)$, $e_y$ falls monotonically.  By extrapolating %%@
to the field at which it vanishes, we may determine $H_{c2}$.  In overdoped cuprates, %%@
fields of $\sim$10 T are enough to go over the ridge.  Figure \ref{Bi2212comp}A shows %%@
the tent profile of $e_y$  in overdoped $\rm Bi_2Sr_2CaCu_2O_{8+y}$ (Bi 2212) revealed %%@
in a field of 30 T.  These curves extrapolate to zero near 50 T which we take to be the %%@
value of $H_{c2}$ at this doping.  The profile in $\rm Bi_2Sr_{2-y}La_yCuO_6$ (Bi 2201) %%@
(Fig. \ref{scale}B) is closely similar in shape.  As $H$ approaches 50 T, $e_y$ %%@
decreases by a factor of 10 to approach zero at 48 T.  

In underdoped hole-type cuprates, however, a field of 30 T is barely sufficient to get %%@
to the top.  Figure \ref{Bi2212comp}B displays curves of $e_y$ in underdoped Bi 2212.  %%@
In comparison with Panel A, the curves in Panel B appear to be more stretched out along %%@
the field axis.  From results on several cuprate families, we have found that this trend %%@
is ubiquitous.  It takes a much larger field to reach the maximum in $e_y$ in underdoped %%@
cuprates (Fig. \ref{scale}A).  To make the trend quantitative, we exploit a scaling %%@
property of $e_y$ vs. $H$ near $T_{c0}$ that we uncovered in Bi-based cuprates.  By %%@
re-plotting the ratio $e_y(H)/e_{y,max}$ versus the reduced field $H/H_{c2}$, we can %%@
collapse curves from samples with different $x$ onto a common curve (Fig. \ref{scale}B).  %%@
Moreover, the similarity applies to curves measured in Bi 2212, 2201 and 2223 (near %%@
their respective $T_{c0}$).  The curve for Bi 2201 at 30 K (Fig. \ref{scale}B) which %%@
extends to 45 T serves as the template against which curves from other Bi-based cuprates %%@
can be compared.

\begin{figure}[h]			% Fig 6
\includegraphics[width=7cm]{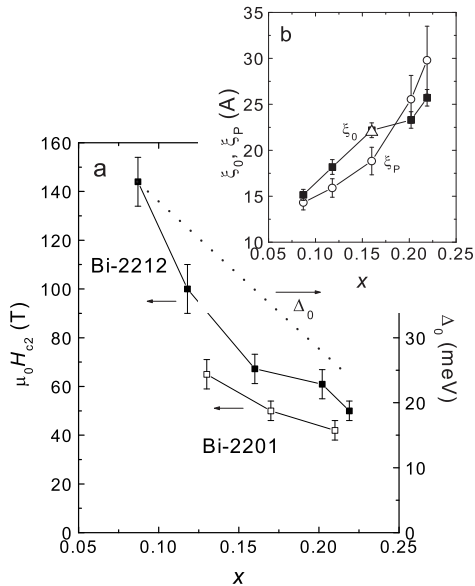}
\caption{\label{Hc2} (Main panel) Variation of $H_{c2}$ with $x$ in Bi 2212 and Bi 2201.  %%@
The dashed line is the ARPES gap amplitude $\Delta_0$ in Bi 2212~\cite{Ding}.  The inset %%@
compares $\xi$ from $H_{c2}$ (solid squares), $\xi_P$ from $\Delta_0$ (open circles) and %%@
$\xi$ from STM spectroscopy (open triangle).  From Ref.~\cite{Wang3}.
}
\end{figure}

Values of $H_{c2}$ derived from the scaling technique are plotted in Fig. \ref{Hc2} for %%@
Bi 2201 and Bi 2212.  The $H_{c2}$ values confirm the qualitative trend inferred from %%@
Fig. \ref{Bi2212comp}.  For Bi 2212, $H_{c2}$ is $\sim$140 T at $x$ = 0.08, and falls %%@
steeply as $x$ increases to 0.25.  The decrease in Bi 2201 is quite similar but the %%@
overall values are smaller.  In LSCO, unfortunately, the large values of $H_m$ make the %%@
scaling technique inapplicable.  However, $H_{c2}$ determined by a different %%@
method~\cite{Wang2} shows a closely similar trend.  

Using the equation $H_{c2} = \phi_0/2\pi\xi^2$, we have computed the coherence length %%@
$\xi$ which is plotted in the inset to Fig. \ref{Hc2}.  At $x$ = 0.08, $\xi$ is small %%@
(1.5 nm), but it steadily increases to 3.0 nm at $x$ = 0.22.  Pan \etal~\cite{Pan} have %%@
measured the decay length of quasiparticle density of states near a vortex core in %%@
optimally doped Bi 2212 by STM and obtained 2.2 nm.  This is in good agreement with our %%@
results (open triangle).  ARPES measurements of the gap amplitude $\Delta_0$ in Bi 2212 %%@
by Harris \etal~\cite{Harris} and Ding \etal~\cite{Ding} show that $\Delta_0$ extends to %%@
$T$ significantly higher than $T_{c0}$ and decreases monotonically with increasing $x$.  %%@
We may use the relation $\xi_P = \hbar v_F/\alpha\Delta_0$ to define the Pippard length %%@
$\xi_P$ (where $v_F$ is the Fermi velocity and $\alpha$ is a number).  Converting the %%@
ARPES gap~\cite{Ding} to $\xi_P$, we find that it agrees with our coherence length if %%@
$\alpha$ is chosen to be $\frac32$ (open circles in inset).  This persuades us that the %%@
3 experiments are measuring the same length scale in Bi 2212.  Hence we should properly %%@
interpret $\Delta_0$ as the superconducting gap amplitude.  Its magnitude dictates the %%@
shortest length scale over which we may bend $\hat{\psi}({\bf r})$, and matches rather %%@
well the vortex core size determined from STM and our $H_{c2}$ measurements.  

From Fig. \ref{Hc2}, we infer that, as $x$ increases from 0.08, the coherence length %%@
which measures the Cooper pair size expands monotonically.  This immediately implies %%@
that the pairing strength starts out being very large in underdoped cuprates, but falls %%@
monotonically with increased doping.  At $x$ = 0.08, we have tightly bound pairs of size %%@
comparable to the interpair spacing.  The sparse density forms a condensate with small %%@
superfluid density $\rho_s$.  Although the onset temperature for pair formation is at %%@
high $T$ (possibly higher than $T_{onset}\sim$ 130 K), the small $\rho_s$ implies low %%@
phase stiffness.  Long-range phase coherence appears at a $T_{c0}$ that is very low.   %%@
As we increase $x$ towards optimal, $\rho_s$ increases rapidly so that long-range phase %%@
coherence appears at a higher $T_{c0}$, but we pay the price of reducing the pairing %%@
strength.  Finally, in the overdoped regime, the rapid decrease of the pairing strength %%@
forces $T_{c0}$ to smaller values despite the large superfluid density available.  The %%@
two conflicting trends appear to account naturally for the dome-shape $T_{c0}$ curve %%@
that is universal in hole-doped cuprates.  The end-point of $H_m(T)$, which is sensitive %%@
to $\rho_s$, determines $T_{c0}$ (see below).  However, the low-temperature onset of %%@
long-range phase coherence determined by $\rho_s$ is emphatically distinct from the high %%@
energy scale of the pairing potential which induces pair formation above 130 K in the %%@
underdoped regime.

\begin{figure}[h]			% Fig 7
\includegraphics[width=6cm]{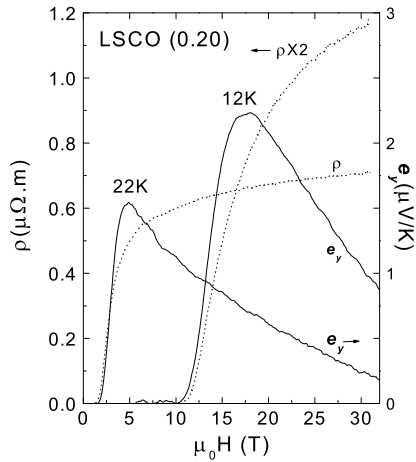}
\caption{\label{fluxflow} The flux-flow resistivity $\rho$ and Nernst signal $e_y$ vs. %%@
$H$ at 22 K and 12 K in LSCO ($x = 0.20$). 
}
\end{figure}

Further clarification derives from a comparison of $e_y$ with the flux-flow resistivity %%@
$\rho$ (Fig. \ref{fluxflow}).  As $H$ exceeds $H_m$ at 22 K, both $e_y$ and $\rho$ rise %%@
nearly vertically.  The field-scale at which $\rho$ forms a knee, often taken to define %%@
`$H_{c2}$', is seen to be just slightly larger than $H^*\simeq$ 5 T (where $e_y$ peaks).  %%@
However, it is quite apparent that the vortex signal remains substantial up to the much %%@
larger value of $H_{c2}\sim$45 T determined by $e_y\rightarrow 0$.  The same discrepancy %%@
is apparent at 12 K.  Flux-flow resistivity can be a rather misleading probe of the %%@
vortex state in cuprates.

Figure \ref{fluxflow} illustrates the difference between the $H$ dependence of $\rho$ in %%@
the vortex liquid state in cuprates and Bardeen-Stephen behavior~\cite{Kim}.  Instead of %%@
a linear increase from zero to the normal-state value $\rho_N(T)$ at $H_{c2}$, $\rho$ %%@
rises steeply by a large fraction ($\sim 0.6$) of $\rho_N$ between $H_m$ and $H^*$, and %%@
then gradually asymptotes to $\rho_N$.  At $H>2H^*$, the vortex liquid is %%@
indistinguishable from the `normal state' using $\rho$ alone.  By contrast, the %%@
difference is apparent in $e_y$.  In the strongly dissipative region above $H_m$ (at low %%@
$T$), long-range phase coherence is absent because of the rapid mobility of the %%@
vortices.  Nonetheless, \emph{local} phase rigidity remains to support a high density of %%@
vortices.  The Nernst signal detects the phase singularity at their cores and allows us %%@
to extrapolate to the field scale at which the vortices are finally suppressed.  

\begin{figure}[h]			% Fig 8
\includegraphics[width=8cm]{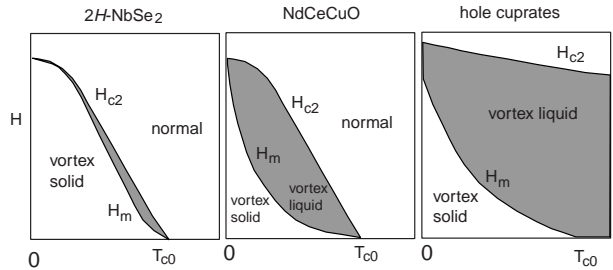}
\caption{\label{vortexcompare}  The fields $H_m$ and $H_{c2}$ in the $H$-$T$ plane in %%@
$2H$-NbSe$_2$, Nd$_{2-x}$Ce$_x$CuO$_4$, and hole-doped cuprates.  In the last (third %%@
panel), $H_{c2}$ is nearly $T$ independent below $T_{c0}$, and the vortex liquid extends %%@
well above $T_{c0}$ .
}
\end{figure}

The loss of phase coherence when $H$ exceeds $H_m$ at low $T$ is closely similar to the %%@
loss of phase coherence when $T$ is increased above $T_{c0}$ in a weak $H$.  In this %%@
light, the initial findings of Xu \etal~\cite{Xu} may be seen as the smooth continuation %%@
of the low-$T$ vortex-liquid state to the $T$ axis above $T_{c0}$.

To place these results in perspective, we compare hole-doped cuprates with the %%@
electron-doped cuprate $\rm Nd_{2-x}Ce_xCuO_4$ (NCCO).  In the latter, the vortex-Nernst %%@
signal rapidly vanishes when $T_{c0}$ is exceeded~\cite{Wang3}.  The $H_{c2}$ line %%@
inferred~\cite{Wang3} from $e_y$ vs. $H$ is just that expected from BCS theory.  The %%@
absence of vortex excitations above $T_{c0}$ in $\rm Nd_{2-x}Ce_xCuO_4$ is likely %%@
related to the absence of a pseudogap state above $T_{c0}$.  Figure \ref{vortexcompare} %%@
compares the $H$-$T$ phase diagrams for $2H$-NbSe$_3$, $\rm Nd_{2-x}Ce_xCuO_4$ and the %%@
hole-doped cuprates.  The first two have a BCS-like phase diagram in which $H_{c2}$ %%@
terminates at $T_{c0}$.  The vortex state is clearly distinguished from the normal state %%@
(the vortex liquid state occupies a much larger area in NCCO).  In hole-doped cuprates, %%@
$H_{c2}$ falls slowly with $T$ (if at all) and seems to approach zero at very high  $T$.  %%@
The vortex liquid state adiabatically continues to $T$ above $T_{c0}$, and no phase %%@
boundary terminating at $T_{c0}$ is observable.  

This viewpoint emphasizes that, in hole-doped cuprates, the Meissner transition at %%@
$T_{c0}$ is invariably the end-point $T_m$ of the melting line.  The zero-$H$ transition %%@
occurs as soon as the population of thermally excited vortex-antivortex pairs are %%@
trapped in the vortex solid phase.  While this picture differs from the BCS scenario, it %%@
is also distinct from what happens at a \emph{strictly} 2D Kosterlitz-Thouless (KT) %%@
transition.  In ${\rm MoGe}$~\cite{Yazdani}, for instance, $T_m$ lies well below %%@
$T_{KT}$.  Interestingly, as we go from $\rm YBa_2Cu_3O_7$ to LSCO to Bi 2212 and Bi %%@
2201 (increasing anisotropy), the $T$ dependence of $H_{m}$ flattens out to approach the %%@
2D KT situation (but, at low enough $H$, $H_m$ always ends at $T_{c0}$).  The $c$-axis %%@
coupling plays a central role in establishing 3D long-range phase coherence.

Finally, we note that, along the classical axis $T$ at $H=0$, there is a large %%@
temperature interval between the mean-field transition scale (perhaps $T^*$) and the %%@
observed $T_{c0}$.  If we scan along the `quantum' axis $H$ at $T=0$, will we find that %%@
the melting field $H_m(0)$ lies significantly below $H_{c2}$?  The nature of the state %%@
in between should be quite unusual.

\end{document}